"Neuropsychological Effects of Rock Steady Boxing in Patients with Parkinson's Disease:
A Comprehensive Analysis"

Lorella Bonaccorsi*, Ugo Santosuosso*, Massimo Gulisano* and Luca Sodini°

April 18 2024

ABSTRACT

This study investigates the efficacy of adapted boxing, specifically Rock Steady Boxing (RSB), in mitigating dopamine decline in individuals with Parkinson's disease. The research involved 40 participants with confirmed diagnosis of Parkinson's diagnosis who underwent biweekly RSB sessions over an 8-week period. Training regimen included activation, core exercises, and a cooldown phase.

The findings revealed a significant amelioration in depressive symptoms through the sessions. Assessment using the Beck Depression Inventory (BDI-II) demonstrated a progressive decrease in scores associated with depressive symptoms, particularly affective, cognitive, and somatic symptoms. The reduction in more severe symptoms was accompanied by an increase in milder symptoms.

Statistical analysis confirmed the significance of the reduction in depressive symptoms over time, suggesting that physical activity, particularly RSB, may contribute to enhancing the quality of life for individuals with Parkinson's disease. The positive impact was observed in both motor and depressive symptoms, suggesting an overall beneficial effect of exercise training.

It is important to note that six participants withdrew from the study due to organizational reasons, resulting in a reduction in the participant count from 40 to 34. Nonetheless, the overall results suggest that RSB could be an effective approach to addressing depression in Parkinson's patients, providing a complementary treatment option to conventional pharmacological therapy.

* Clinical and Experimental Medicine Department, University of Study of Florence, Largo Brambilla, 3 - 50134 Firenze, Italy
° University of Study of Milan , Milan, Italy

lorella.bonaccorsi@unifi.it, ugo.santosuosso@unifi.it, massimo.gulisano@unifi.it, luca.sodini@studenti.unimi.it.

## Introduction

Parkinson's disease (PD) is a neurodegenerative disorder that affects the Basal Ganglia (BG), which are responsible for controlling voluntary and automatic movements (1). The disease is caused by the loss of cells that produce Dopamine, a neurotransmitter that regulates muscle movements. Its onset typically occurs during an individual's sixth decade of life (2).

The BG are divided into two systems: the pyramidal system, which controls voluntary movement, and the extrapyramidal system, which regulates automatic movement. The extrapyramidal system is responsible for regulating tone and posture, controlling axial and proximal muscles of the limbs, and inhibiting involuntary movements. The extrapyramidal system is composed of the Basal Ganglia (BG), which are further divided into the neostriatum, paleostriatum, black matter, subthalamic nucleus of Luys, and the Bucy circuit (3).

The nigrostriatal pathway is the primary pathway involved in PD. It is a dopaminergic pathway that connects the Substantia Nigra pars compacta to the corpus striatum. The corpus striatum receives glutamatergic afferents from the cortex and dopaminergic afferents from the Substantia Nigra pars compacta, forming the nigrostriatal pathway. Its efferent neurons are GABAergic and thus inhibitory. The neuronal population of the corpus striatum can be divided based on the dopaminergic receptors it expresses: those expressing the D1 type are excited by dopaminergic stimulation from the substantia nigra, while those expressing the D2 type are inhibited by that pathway. These two distinct populations of striatal neurons project to different areas, giving rise to the direct and indirect pathways, both terminating at the Ventral Anterior (VA) and Ventral Lateral (VL) nuclei of the thalamus, underscoring the significance of the nigrostriatal dopaminergic pathway in movement control.

The direct pathway and the indirect pathway are two neural pathways within the basal ganglia that play opposing roles in movement modulation. The direct pathway initiates and facilitates movement, while the indirect pathway is involved in inhibiting or controlling movement. In PD, the predominantly presynaptic dopamine deficit caused by degeneration of the main production center of this neurotransmitter, namely the Substantia Nigra pars compacta, is the reason why the indirect pathway prevails over the direct pathway. This results in hyperactivity of the inhibitory outputs to the thalamus and cortex, leading to the generation of symptomatic manifestations.

The incidence of PD is 8-18 cases per 100,000 person-years, and the prevalence is 80-130 sufferers per 100,000 people (3). It is less frequent in China, Japan, and in the black and Asian population. The ratio of affected males:females is 1-2:1. Some studies have found no significant differences between the genders (4). The preclinical phase of Parkinson's disease can persist for 10 to 20 years, followed by the onset of symptoms, which typically occurs around the sixth decade of life. Subsequently, living conditions generally remain manageable for about 20 years, after which significant disability may develop (4).

Most individuals with PD have an idiopathic condition, meaning it has no known specific cause. However, a small percentage of cases can be associated to known genetic factors. Other factors have been linked to the risk of developing the disease, but no causal relationships have been established. The disease occurs sporadically in 80-85% of cases, while in 15-20% of cases, there is a familial predisposition. Familial predisposition to PD is linked to the identification of several genes that are associated with specific forms of the disease (5). There are a small number of genes that are involved into 5-6% of total PD and there are other genes that lead to an increase risk of PD. Approximately 15% of PD subjects have a familiar member who has also developed PD. 5–10%

PD cases can be attributed to monogenic forms. Other factors are a combination of complex genetic susceptibility and environmental causes. For the monogenic forms, there are several genes, related to both an autosomal dominant (SNCA, LRRK2, and VPS35) and an autosomal recessive (PRKN, PINK1, DJ1) inheritance. Moreover, there is X-linked inheritance and complex parkinsonian phenotypes due to mutations in the ATP13A2, DCTN1, DNAJC6, FBXO7, PLA2G6, and SYNJ1 genes. Furthermore, other genes, such as CHCHD2, LRP10, TMEM230, UQCRC1, and VPS13C have been recently involved in PD. Finally, mutations in genes such as glucocerebrosidase (GBA) might play a role both as a monogenic cause and a genetic susceptibility factor (6). Homozygosity for this mutation characterizes Gaucher syndrome, which involves a deficiency in sphingolipid metabolism (7).

In PD, there is a reduction and depigmentation of dopaminergic neurons in the Substantia Nigra pars compacta of the midbrain, along with the presence of extracellular melanin and intracellular Lewy bodies (8). These Lewy bodies are aggregates of alpha-synuclein and ubiquitin, present in 80% of PD cases. Additionally, there is some degree of reactive gliosis and minimal neuronal loss at other loci. However, the proposed staging based on Lewy bodies has not been utilized to assess disease progression. Lewy bodies are found in various areas of the brain, such as the olfactory bulb, dorsal nucleus of the vagus, and trunk nuclei. In the early phase of the disease, Lewy bodies mediate early non-motor symptoms such as hyposmia, sleep disturbances, constipation, anxiety, and depression. In the intermediate phase, they are observed in the Substantia Nigra pars compacta, whose degeneration causes the motor symptoms. In the late stage of PD, they are detected at the cortical level, leading to cognitive and clinical symptoms similar to DLBD, a neurodegenerative disease characterized by the widespread presence of Lewy bodies. However, this classification has not been adopted to assess the progression of PD, unlike the widely used Hoehn and Yahr scale, which is employed to observe and measure the symptomatic progression of PD (9-11).

The pathophysiology of PD involves the Basal Ganglia (BG), which consist of two regions: one that primarily receives input from the cortex and another that transmits GABAergic, hence inhibitory, outputs to the thalamus. From there, thalamic projections extend to the cortex, particularly to the frontal lobes, in areas primarily involved in the planning and execution of purposeful movements. The latter region is comprised of the internal globus pallidus, or GPi, and the substantia nigra pars reticulata, or SNpr. Additionally, the external globus pallidus, or GPe, and the subthalamic nucleus of Luys, or STN, are part of the BG. These two regions, one receiving input from the cortex and the other sending output to the cortex, are interconnected by two pathways. The direct pathway extends from the striatum to the inner segment of the globus pallidus (GPi) and to the substantia nigra pars reticulata (SNr), transmitting inhibitory GABAergic outputs. Dopamine, acting on D1 receptors in the striatum, facilitates this pathway, promoting movement. The indirect pathway involves inhibitory GABAergic fibers from the striatum to the outer segment of the globus pallidum (GPe) and from there to the subthalamic nucleus (STN), which in turn has an excitatory action on the inner globus pallidus (GPi). Dopamine, by binding to D2 receptors in the striatum, inhibits this pathway, facilitating movement. In PD, the predominantly presynaptic dopamine deficit caused by degeneration of the main production center of this neurotransmitter, namely the Substantia Nigra pars compacta, is the reason why the indirect pathway prevails over the direct pathway. This results in hyperactivity of the inhibitory outputs to the thalamus and cortex, leading to the generation of symptomatic manifestations (12-13).

The subclinical phase (10-15 years before onset) of PD is characterized by the involvement of the circuit between the limbic system and the autonomic nervous system (ANS), leading to hyposmia, constipation, sleep disturbances, and restless legs syndrome. During this phase, neuronal loss is less than 70% and there is sympathetic system compensation. In the clinical phase, there are motor and nonmotor symptoms, with 70% to 90% neuronal loss, and good drug sensitivity. In the advanced stage, there is greater than 90% neuronal loss and pharmacological resistance.

The classic triad of Parkinson's disease symptoms consists of resting tremor, bradykinesia, and rigidity. Dyskinesias refer to involuntary and abnormal movements, which can be a side effect of long-term treatment with levodopa, a drug commonly used to manage Parkinson's disease symptoms. As for tremor, it arises at rest as an initial symptom of the disease. It is present in only 70% of cases, occurs asymmetrically, with frequency 4-6 Hz, typically involving the extremities of the upper limbs. It tends to appear during rest or under emotional stress, then disappears during activity and in sleep (14-15).

Regarding bradykinesia, it is characterized by slowness of movements, particularly in rapid and alternating movements, which is manifested at the level of the facial muscles (hypomimia), the hand (micrographia), the limbs (reduced pendular synkinesias during walking), the muscles of swallowing (less swallowing causing sialorrhea), the muscles involved in phonation (unilateral hypophonic dysarthria), and low walking (16).

The non-motor symptoms are present in approximately 40% of Parkinson's disease cases. These symptoms are associated with impaired signal transmission systems of dopaminergic, noradrenergic, and serotonergic nerve stimuli. The presence of these symptoms is also frequently linked with anxiety disorders (17-18). Apathy is a distinctive and independent symptom due to a deficit in frontal subcortical circuits in PD (19). Regarding cognitive disorders, there is a dysexecutive syndrome that can evolve to frank dementia in advanced stages. In psychiatric disorders, there are angry outbursts, critical visual hallucinations, often triggered by presence or sudden withdrawal of therapeutic treatments or infections. (20-21).

## Materials and Methods

Study setting

The questionnaire used is the Beck Depression Inventory (22), a self-report questionnaire used in the literature to detect depressive symptoms (and not to diagnose depression) as described in the Diagnostic and Statistical Manual of Mental Disorders (DSM) (23-25). This version consists of 21 questions, most of which have five possible responses presented on a scale of 0 to 3, with some of these having intermediate options (for example, there are responses 2 and 2b). The score obtained by summing the scores of individual questions indicates the presence and, if any, severity of depressive symptoms: 0-10 points indicate a "depression level compatible with normality"; 11-14 points indicate "a state of dysphoria on the border with pathological aspects"; 15-16 points indicate a "dysphoric state that causes discomfort and difficulty for the person"; above 16 points indicate a "particularly difficult situation and an index of depressive reaction that in some cases may be particularly severe". The aforementioned scores apply to a questionnaire administered to a man, while for a questionnaire administered to a woman, the scales are slightly different: the first scale goes from 0 to 13, the second from 14 to 17, the third from 18 to 19 and the fourth with at least 20 points. The questions can be divided into two subgroups based on the nature of the questions: Cognitive and Somatic-affective. The qualities of the BDI as a tool for measuring the psycho-affective response of the subjects under examination are: ease of use and interpretation of results, the possibility of self-administering the test, since some of the proposed questions are highly personal, and the study is based on the anonymity of the participants (26).

Study design

In the group attending the boxing course, 40 subjects with a confirmed diagnosis of PD were initially examined, 75% of whom were men and 25% women, with an age range of 51 to 85 years and an average age of 68 years.

Data collection took place in three separate periods during which the questionnaire was administered: the first period was from May 2 to May 8, 2023; the second from May 29 to June 9, 2023, and the third period from June 26 to July 20, 2023.

In the first period, participant recruitment and the first administration of the questionnaire took place. Immediately after the training session, the subject was accommodated in a room where they found the BDI and the privacy notice to sign. The patients gave their informed consent to participate in the study. Initially, each patient was assigned a unique alphanumeric code. The purpose of this code is to maintain the anonymity of the participants while allowing the questionnaires of the same person to be reunited after the end of data collection.

After writing the code and receiving instructions on how to use the questionnaire, the participant was left alone in the room until they finished answering the questions, after which the BDI was inserted into an envelope along with the other questionnaires. The same procedure was carried out in the two subsequent data collection periods.

Between the first and second periods, the participants attended the biweekly boxing sessions, so on average, the subjects underwent 8 workouts, and the same occurred between the second and third periods of the study A total of 16 to 20 workouts were carried out between the first and last questionnaire. During the third data collection period, 6 people (5 men and 1 woman) dropped out of the study for organizational reasons, reducing the number of participants in the study from 40 to 34 people.

Statistical analysis

The preliminary analyses were conducted using the Statistical Package for Social Sciences (SPSS, Evanston, IL, USA) and the jamovi software (The jamovi project 2023; jamovi (Version 2.3), Computer Software, retrieved from https://www.jamovi.org). Preliminary analyses were conducted using these software to examine anomalous, missing, and incomplete data, as well as to perform normality tests on the distribution. As an exclusion criterion, subjects who did not provide responses or had fewer than 50% of the total number of responses were removed. Results were expressed as median or mean ± standard error, if not specified otherwise.

Statistical analysis of continuous variables was carried out using the paired t-test. This test is used to compare the means of two groups in which observations are paired or matched. The paired t-test evaluates whether there are significant differences between the means of the two distributions. It is important to note that the paired t-test compares differences between paired observations, not the direct means of the two groups. Additionally, the test requires that the differences be normally distributed.

The presumed associations between ordinal and nominal variables were explored using the chi-square test ($\chi^2$). This statistical test was used to evaluate whether there is a significant association between two categorical variables. In particular, the test compares the observed frequency values in a contingency table with the expected values that would be observed if the two variables were not correlated. In the context of experimental analysis, univariate and multivariate associations and comparisons were conducted using the repeated measures ANOVA test. This test was used to evaluate the effects of boxing treatment on the variable "depression" in the study group over time. In particular, the repeated measures ANOVA is used to analyze data in which measurements are taken on the same subjects at different times, in order to examine whether there are significant differences between the means of multiple groups based on intra-subject variability.

The multivariate ANOVA can be used when there are more than two variables and allows for the estimation of the contribution of qualitative independent variables that include both ordinal (such as questionnaire scores) and nominal (such as gender) variables. ANOVA compares the calculated variance between groups with the variance within groups. Due to the parametric distribution, the data were analyzed as paired samples, and correlations between variables were determined using the Pearson correlation test and the Kendall correlation coefficient. The p-value, when less than 0.05, was considered the threshold for accepting the significance of the test ($p<0.05$ = statistically significant) for ANOVA, t-test and chi-square tests.

Correlation is a relationship between two variables based on a certain regularity, and it does not depend on a cause-effect relationship, but rather on the tendency of one variable to change based on the other. The Pearson correlation index, also known as the linear correlation coefficient, expresses a possible linear relationship between two statistical variables. This index is used to evaluate the strength and direction of a linear relationship between two continuous variables.

The Spearman correlation index, also known as the Spearman rank correlation coefficient, is a non-parametric statistical measure of correlation. It measures the degree of relationship between two variables and requires only the assumption that they be orderable, and, if possible, continuous. Unlike the Pearson linear correlation coefficient, the Spearman correlation coefficient does not measure a linear relationship even when interval measures are used. Instead, it allows for the establishment of how well a relationship between two variables can be described using a monotonic function, that is, one that preserves the order between values.

Results

Depression is one of the most debilitating non-motor symptoms of Parkinson's disease (PD) as it deteriorates the quality of life for both subjects with PD and their families. Physical exercise is often prescribed as an additional treatment associated with conventional therapy to slow down the progression of symptoms. Improving depressive symptoms, along with stabilizing motor symptoms, can have additional benefits for a better quality of life of affected subjects. This study aims to evaluate the effect of physical activity, particularly Rock Steady Boxing (RSB), on the severity of depressive symptoms. The Beck Depression Inventory-II (BDI-II) is used to assess depression symptoms psychometrically and diagnostically.

The BDI-II is the standard reference tool for measuring the severity of depressive symptoms in Parkinson's and other associated conditions. It aligns with the diagnostic criteria outlined in the Diagnostic and Statistical Manual of Mental Disorders (DSM-IV) by the American Psychiatric Association, which identifies the broad nature of depressive symptoms divided into cognitive, affective, and somatic categories. The model used in this study is structured around these three factors. The depression severity threshold scores used in this study distinguish four levels ranging from minimal depression (0-13) to mild (14-19), moderate (20-28), and severe (29 and above).

Data collection was conducted by the same blinded expert evaluator as described in the methods. Questionnaires were administered to participants at the beginning of the study (time 0), after 1 month, and after 2 months to analyze the effects of physical exercise during the observation period. The impact of the presence or absence of pharmacological therapy was not considered in this analysis. Since a predefined boxing protocol was not available, an exercise protocol designed by a certified coach was adapted for the study. RSB exercises were performed under the coach's guidance. The lessons were held twice a week for 8 weeks, each lasting 60 minutes, including physical and mental activation exercises (10 minutes), discipline-specific exercises (40 minutes), and cool-down exercises (10 minutes). The primary goal of this study was to assess the long-term effect of RSB activity on the severity of depression symptoms, which is associated with both disease severity and the perception of symptom impact on daily life management. The study was conducted on a sample of 40 participants with confirmed Parkinson's diagnosis for over 3 months, including 30 males and 10 females. The average age was 67.9±10.1 years (range 51-85) for males and 68.3±9.5 years (range 56-84) for females (Figure 1).

|  | Sesso | Età | Assente |
|---|---|---|---|
| N | M | 30 | 30 |
|  | F | 10 | 10 |
| Media | M | 67.900 | 0.000 |
|  | F | 68.300 | 0.000 |
| Deviazione standard | M | 10.080 | 0.000 |
|  | F | 9.534 | 0.000 |
| Minimo | M | 51 | 0 |
|  | F | 56 | 0 |
| Massimo | M | 85 | 0 |
|  | F | 84 | 0 |

FIGURE 1

Figure 1 shows a descriptive analysis of dataset. Forty individuals with a confirmed diagnosis of Parkinson's disease for over 3 months were analyzed. The cohort consisted of 30 males and 10 females. The average age ± standard deviation was 67.9±10.1 years (range 51-85) for males and 68.3±9.5 years (range 56-84) for females. The questionnaire was completed by 34 participants (85%). During the questionnaire administration, 6 participants, including 5 males and 1 female, did not complete the evaluation cycle for reasons unrelated to the program. These participants were excluded from the analysis. The questionnaire was completed by 34 participants (85%). During the questionnaire administration, 6 participants, including 5 males and 1 female, did not complete the evaluation cycle for reasons unrelated to the program. These participants were excluded from the analysis.

In Figure 2 we describe, on the left, the box plot (box-and-whisker plot) of the demographics of the sample categorized by gender of the subjects. In the box plot, the mean of the age variable is represented by a thick black line for the male and female categories, while the colored area represents the standard deviation (SD) of the samples for the two categories. The points represent cases outside the SD (outliers). On the right, the histograms show the frequency of cases based on age, i.e., the distribution of the age variable, separated by gender.

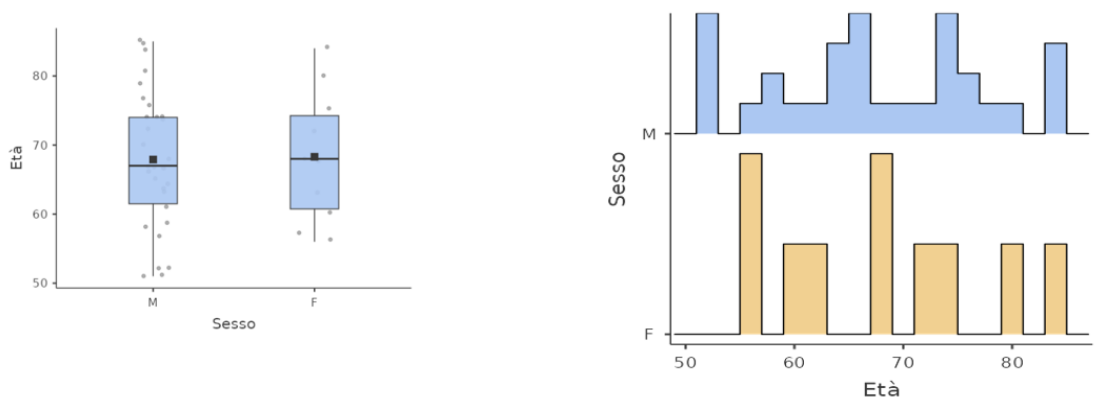

FIGURE 2

Figure 2 describes the demographic details of the sample and shows the box plot of the subjects on the left, with the median age represented by a thick black line for both male and female categories, while the colored area represents the standard deviation of the participants for the two categories. The points represent cases outside the standard deviation (outliers). On the right, the histogram describes the frequency of cases based on age, i.e., the distribution of the age variable, divided by gender.

The prevalence of PD is constantly increasing due to the aging of the population, with an estimated 220,000 patients in Italy. Although it is rarely the primary cause of death, subjects with PD have a lower life expectancy and a higher mortality rate due to frequent infections and falls.

The progression of the disease is characterized by a high degree of variability, The demographic information of the sample is essential for understanding the characteristics of the participants. This information is useful for planning and conducting studies on Parkinson's disease.

The analysis began with an analysis of the distribution of the values assigned to each response, divided by type (affective, cognitive, and somatic) and with the total count of missing responses for the first, second, and third session (Figure 3).

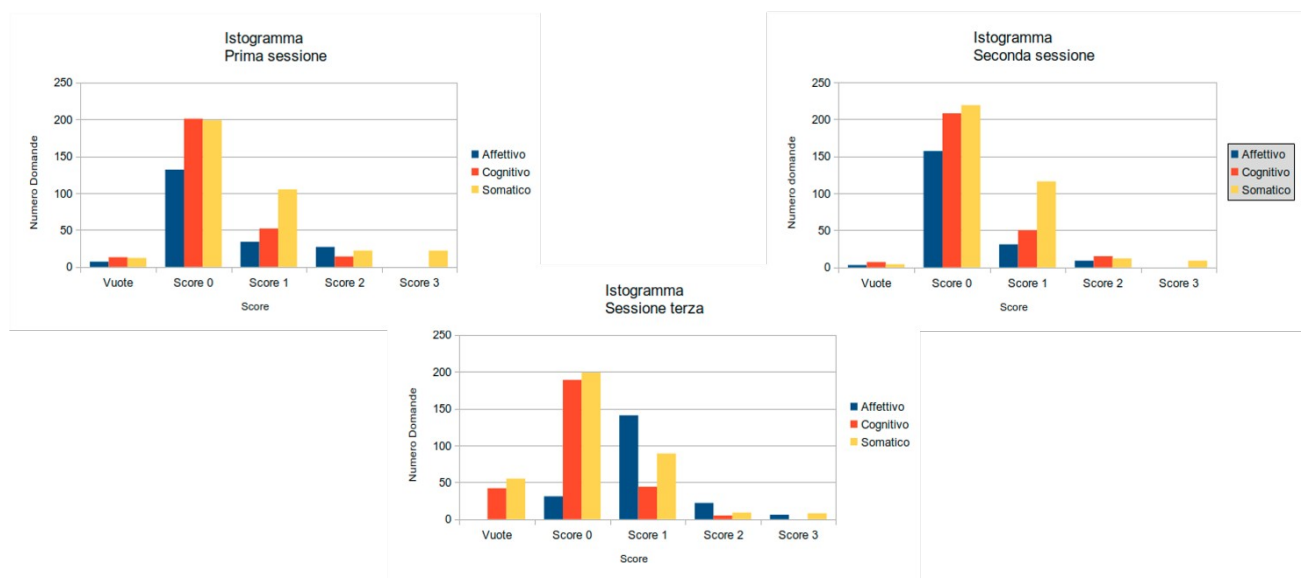

FIGURE 3

The figure 3 summarizes the distribution of the values of the scores assigned to each response, divided by type (affective, cognitive, and somatic), and the total of missing responses, respectively, for the first, second, and third session.

The results display that in all sessions, responses with a score of 0 and 1 are predominant, and their trend over time tends to increase. On the other hand, a decreasing number of responses with higher scores are observed in subsequent sessions. In Figure 4, the distribution of the questionnaire results with respect to the categories associated with increasing levels of depressive symptoms over time is reported. The trend of symptoms is consistent with the results described above. The "raw" values are obtained by summing the scores assigned to each question, while the "normalized" values are calculated by estimating the empty responses present in each questionnaire. The data highlights that the obtained responses falls between the absence and the presence of mild depressive symptoms, which decrease in subsequent sessions (Figure 4).

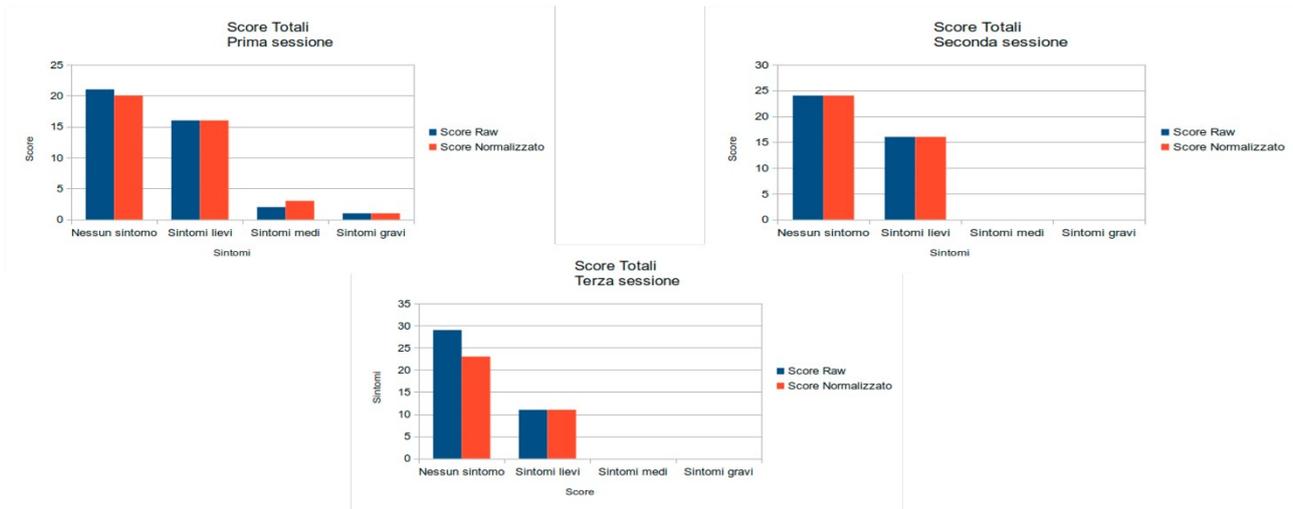

FIGURE 4

Figure 4 describes the distributions of the questionnaire results associated with the depression symptom scores in the first, second, and third session. The "raw" values are obtained by summing the scores assigned to each question, while the "normalized" values are calculated by considering the empty responses present in each questionnaire

Subsequently, we explored the set of collected responses concerning each score in the different sessions (Figure 5).

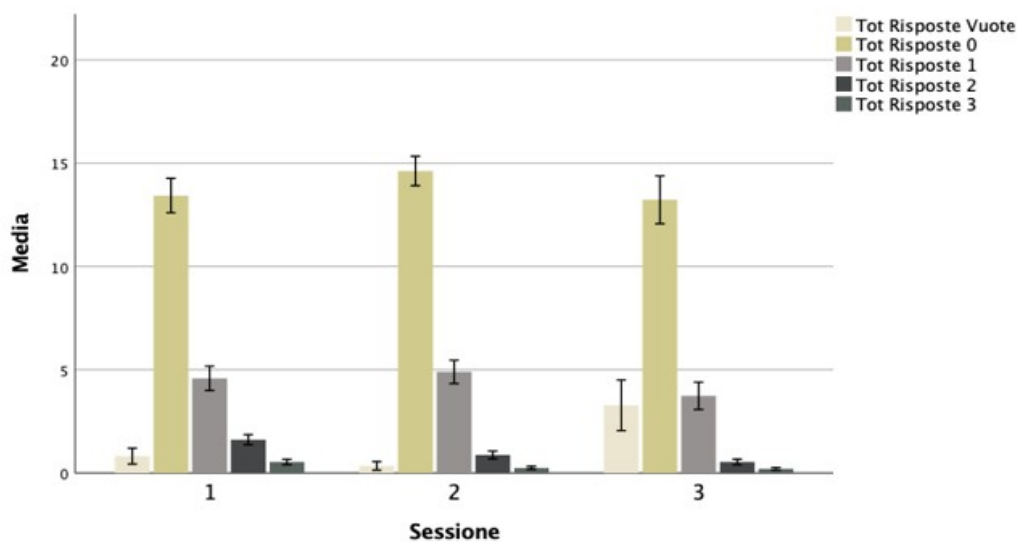

FIGURE 5

Figure 5 describes the average scores for each type of response (from 0 to 3) are represented in the histograms, considering all the questionnaires. The hypothesis testing using the chi-square distribution "$\chi^2$" has confirmed a significantly different distribution in subsequent sessions for empty responses, for those with a score of 0, and for responses with a score of 2. The average values are in x axis versus the sessions reported in y axis.

The average score for each type of response (from 0 to 3) is represented in the histograms, considering all the questionnaires. The hypothesis testing using the chi-square distribution "$\chi^2$" has confirmed a significantly different distribution in subsequent sessions for empty responses, for those with a score of 0, and for responses with a score of 2. The distribution displays a tendency to decrease over time for responses with higher depression scores. We then calculated the counts of the questionnaire responses in the three categories of increasing depressive symptoms (Figure 6). The results show a reduction, although not significant, in more severe symptoms and a corresponding increase in less severe symptoms in the three subsequent sessions (Figure 6).

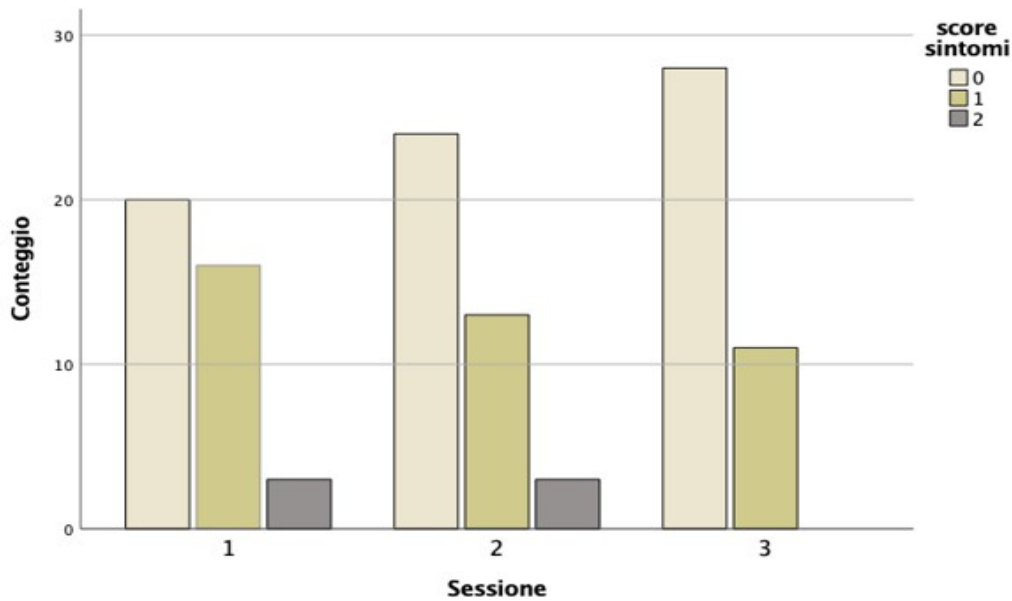

FIGURE 6

The figure 6 shows the total counts of the scores of the results from all questionnaires (y axis) in the following sessions (x axis), dividing them into three categories from absence of depression symptoms to severe symptoms (from 0 to 2).

The previous result is confirmed by the average score of all responses categorized by symptoms in the sessions, which shows a trend of reducing the severity of symptoms over time, although not significant (Figure 7).

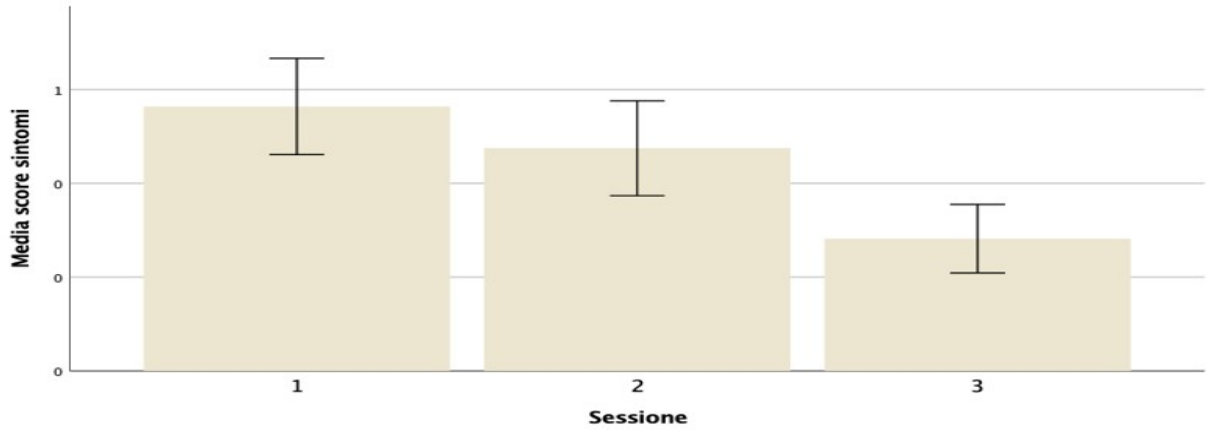

FIGURE 7

Figure 7 reports the average values of the symptom scores (from a value of 0 to a value of 2) from all questionnaires (x axis) in the three following sessions (y axis).

Subsequently, we counted the scores of the responses for each of the three factors (affective, cognitive, and somatic). The results show a trend towards a decrease in responses with higher scores and a reciprocal trend towards an increase in lower scores over time, with a statistically significant distribution in different sessions responses only for the affective factor (Figure 8).

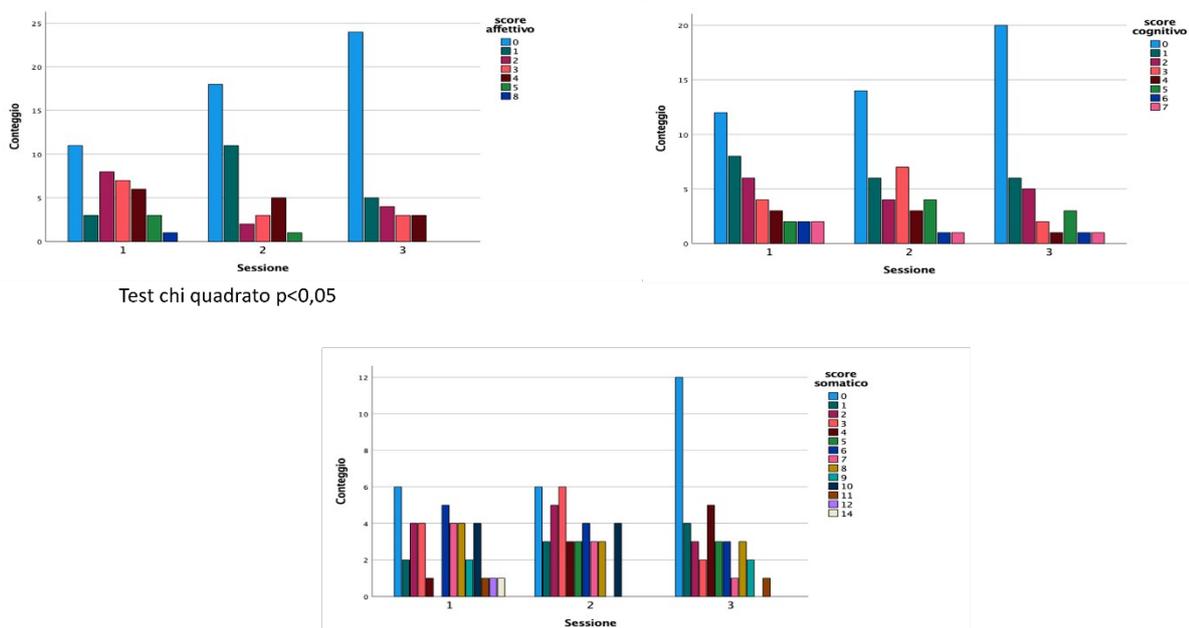

FIGURE 8

Figure 8 presents that the counts of the scores of each response for the three factors (affective, cognitive, and somatic) in x axis in the analyzed sessions in y axis. The distribution of responses in subsequent sessions is significantly different only for the affective item.

In Figure 9, the average scores of individual responses (empty, 0, 1, 2, 3) for the affective, cognitive, and somatic factors are described. For the affective and cognitive factors, the values are distributed significantly differently across sessions with a reduction in responses with higher scores. For somatic item, the distribution of average values is not significantly different.

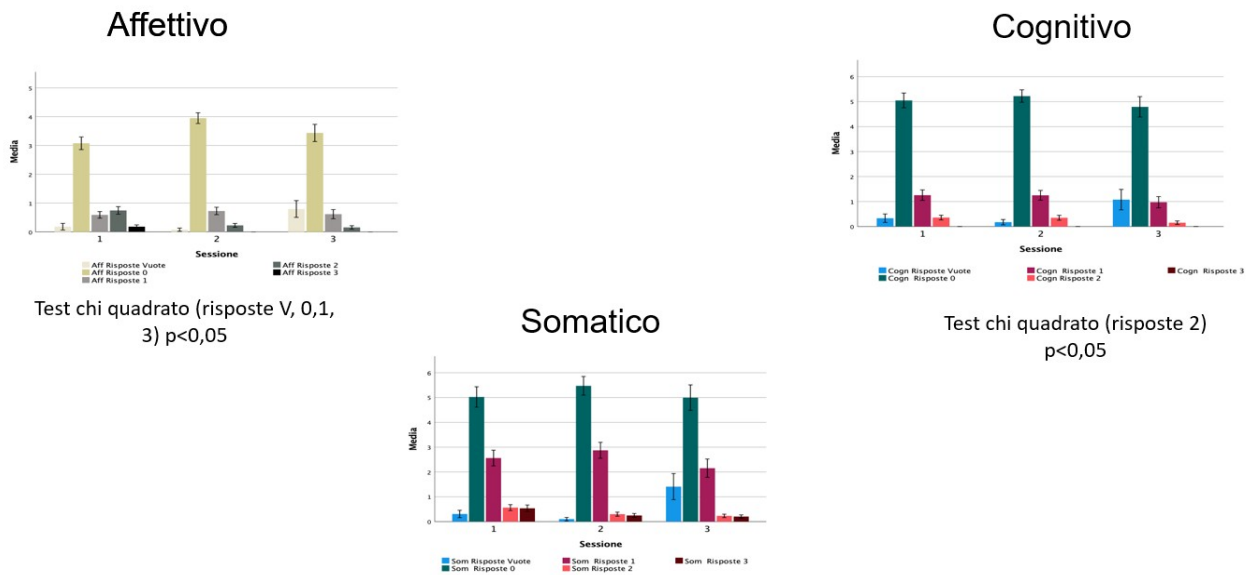

FIGURE 9

In figure 9 the average values of scores of individual responses (empty, 0, 1, 2, 3) for the affective, cognitive, and somatic factors. For the affective and cognitive factors, the values are distributed significantly differently across different sessions. For the somatic factor, the distribution of average values is not statistically significant.

Subsequently, we performed a repeated measures ANOVA analysis on the symptom scores for all items (Figure 10-11).

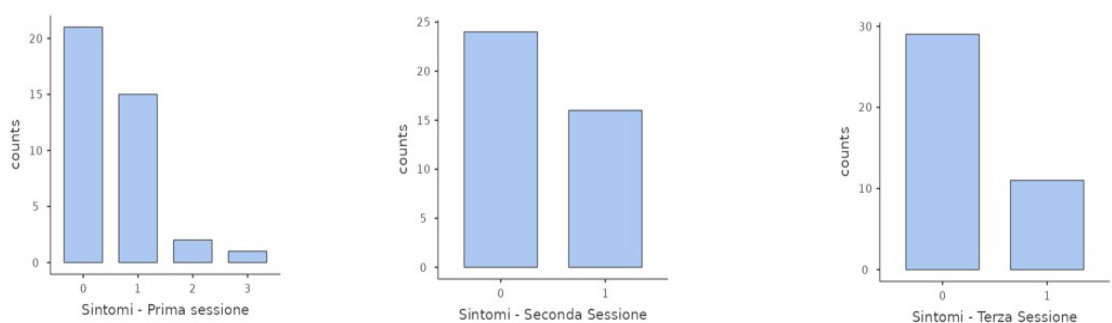

FIGURE 10

The figure 10 highlights the repeated measures ANOVA analysis on the symptom counts (from 0 to 3) in the three subsequent questionnaire sessions. On the x-axis, 0 corresponds to no symptoms; 1 corresponds to mild symptoms, 2 to moderate symptoms, and 3 to severe symptoms of PD.

| Effetti Entro i Sggetti | | | | | | |
|---|---|---|---|---|---|---|
| | Somma dei Quadrati | gdl | Media Quadratica | F | p | η²G |
| MR Fattore 1 | 336.617 | 2 | 168.308 | 12.007 | <.001 | 0.071 |
| Residuo | 1093.383 | 78 | 14.018 | | | |

Nota. Somma dei quadrati Tipo 3

FIGURE 11

The figure 11 describes the results of the repeated measures ANOVA analysis of symptom counts across sessions. As can be observed from the p value < 0.001, the difference in scores obtained between sessions is statistically significant.

Figure 10 shows the plot of the ANOVA analysis on symptom counts (from 0 to 3) obtained from questionnaires in the three sessions. Figure 11 illustrates the results of the repeated measures ANOVA analysis of symptom counts across sessions. The p value is less than 0.001, indicating a highly significant difference in scores between sessions. Regarding the study of symptoms, the results show that the difference between scores in subsequent sessions is statistically significant (Figure 10-11). We repeated the investigation on individual factor scores. For the affective factor, a statistically significant decrease in scores of affective-type questions associated with depressive state over time is detected, and it is present already in the second session and persists in the third.

Figure 12 repots the trend of scores of affective-type questions categorized by gender over time; the colored dots are the real values, the white dots represent the averages, and the vertical lines indicate the standard deviations.

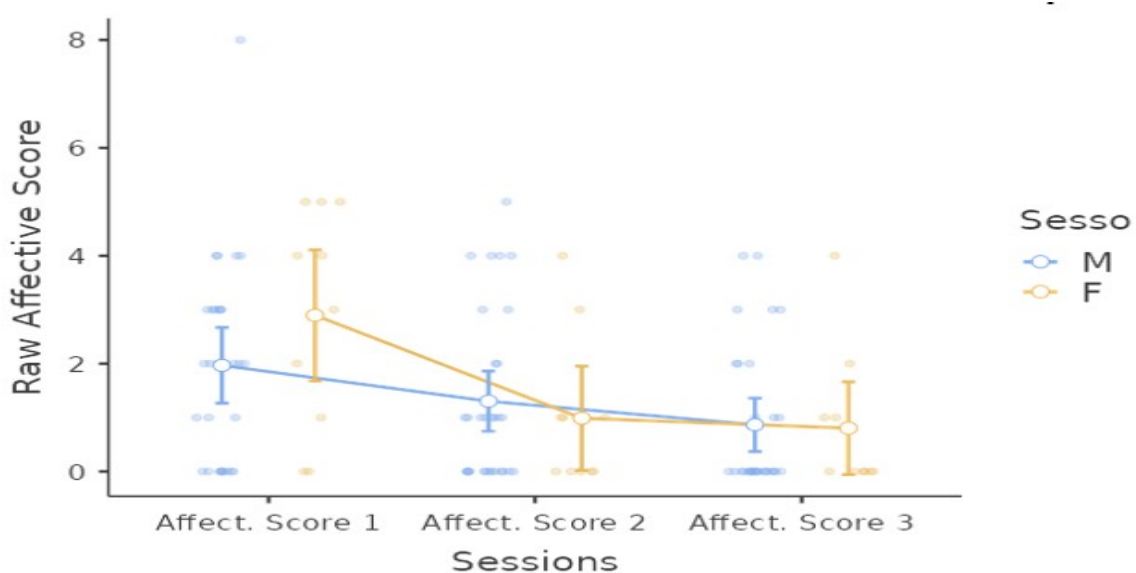

FIGURE 12

The plot in Figure 12 represents the repeated measures ANOVA analysis of the affective item over time. It shows the trend of scores of affective-type questions categorized by gender (M: male/F: female); the colored dots are the real values, the white dots represent the mean for different scores, and the vertical lines indicate the standard deviation.

Figure 13 displays the results of the ANOVA analysis on the calculated scores of affective factor responses from the questionnaires without considering the effect of confounding variables. The p-value (less than 0.001) indicates that the difference in scores for the affective item between sessions is highly significant.

Effetti Entro i Sggetti

|  | Somma dei Quadrati | gdl | Media Quadratica | F | p | η²G |
| --- | --- | --- | --- | --- | --- | --- |
| Sessions | 38.850 | 2 | 19.425 | 14.053 | <.001 | 0.115 |
| Residuo | 107.817 | 78 | 1.382 |  |  |  |

Nota. Somma dei quadrati Tipo 3

FIGURE 13

The results of the scores of affective item responses of the questionnaires without further confounding variables are described in figure 13. The difference in scores for the affective item among the sessions is statistically significant (p-value less than 0.001).

Later, we evaluated the effect of the covariates "age" and "gender", taken separately, on the scores of affective-type responses for the determination of depressive state before, during, and after physical activity treatment. The analysis does not detect statistically significant variations. A possible explanation could be the excessive fragmentation of the sample, which disperses the information of the data with respect to the number of cases (Figure 14).

Effetti Entro i Sggetti

|  | Somma dei Quadrati | gdl | Media Quadratica | F | p | η²G |
| --- | --- | --- | --- | --- | --- | --- |
| Sessions | 4.824 | 2 | 2.412 | 1.855 | 0.164 | 0.017 |
| Sessions ✱ Età | 5.127 | 2 | 2.564 | 1.971 | 0.147 | 0.018 |
| Sessions ✱ Sesso | 6.457 | 2 | 3.228 | 2.482 | 0.091 | 0.022 |
| Residuo | 96.251 | 74 | 1.301 |  |  |  |

Nota. Somma dei quadrati Tipo 3

FIGURE 14

The figure 14 shows the results of the scores of affective item answers. Data with and without the effects of the covariates "age" and "gender" were examined separately. As can be observed from the p-values, there are no statistically significant variations, and each of the variables analyzed is unable to explain the variability of the model.

Regarding the cognitive item, the trend of scores for answers categorized by gender (M: males, F: females) in the different sessions a reduction over time was showed in Figure 15.

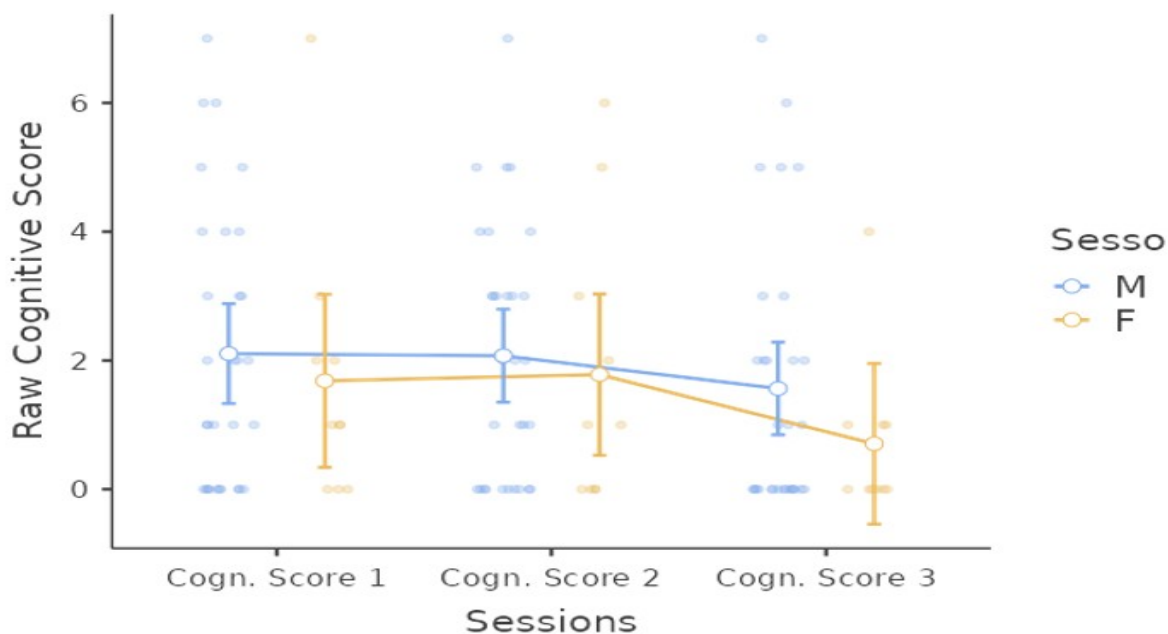

FIGURE 15

The above plot depicts the trend over different time points of the scores for cognitive factor answers categorized by gender in the three sessions; the colored points represent the distribution of actual values; the white dots indicate the mean values, and the vertical lines show the trend of standard deviations. The difference in scores for the cognitive item among the sessions is also statistically significant (p-value less than 0.031) (Figure 16).

Effetti Entro i Sggetti

| | Somma dei Quadrati | gdl | Media Quadratica | F | p | $\eta^2_G$ |
|---|---|---|---|---|---|---|
| Sessions | 11.267 | 2 | 5.633 | 3.639 | 0.031 | 0.023 |
| Residuo | 120.733 | 78 | 1.548 | | | |

Nota. Somma dei quadrati Tipo 3

FIGURE 16

Figure 16 describes the ANOVA analysis on the scores of cognitive item obtained from the questionnaires without further confounding variables. The difference in scores for the cognitive item in the different sessions is statistically significant (p-value less than 0.031).

The scores reduction trend of these questions is more evident in the last session, suggesting that detecting an improvement in depression through cognitive-type responses may require longer periods of physical activity administration. The ANOVA analysis on cognitive item answers investigating the effects of the covariates "age" and "gender" taken separately, highlights a statistically significant variation in the scores of the sessions, adding the covariate "age" ("p value" equal to 0.004), but not for the covariate "gender". This result is probably due to an imbalance in the number of males and females in the sample (Figure 17).

Effetti Entro i Sggetti

| | Somma dei Quadrati | gdl | Media Quadratica | F | p | η²G |
|---|---|---|---|---|---|---|
| Sessions | 11.974 | 2 | 5.987 | 4.319 | 0.017 | 0.026 |
| Sessions * Età | 16.666 | 2 | 8.333 | 6.011 | 0.004 | 0.036 |
| Sessions * Sesso | 1.317 | 2 | 0.659 | 0.475 | 0.624 | 0.003 |
| Residuo | 102.579 | 74 | 1.386 | | | |

*Nota.* Somma dei quadrati Tipo 3

FIGURE 17

The Figure 17 describes the ANOVA test on the scores of cognitive item, assuming that all cases were analyzed across sessions together. Subsequently, we studied the effect of the "age" and "gender" covariates, that were examined separately over time. As indicated by the "p-values", the difference in the scores of the different sessions is statistically significant concerning sessions overall and age, while it is not significant for sessions and gender covariates.

The ANOVA test on the scores of somatic item answers was performed, assuming that data were analyzed in the sessions together regardless of other variables The reduction in scores for the somatic item is statistically significant in the different sessions without other covariates (p-value < 0.001) and the decrease in scores obtained from answers follows a constant trend over time (Figures 18-19). Then we studied the effect of the "age" and "gender" covariates examined separately. As indicated by the p-values, the difference in scores of the specific sessions is statistically significant concerning sessions and age variables, while it is not significant for sessions and gender variables.

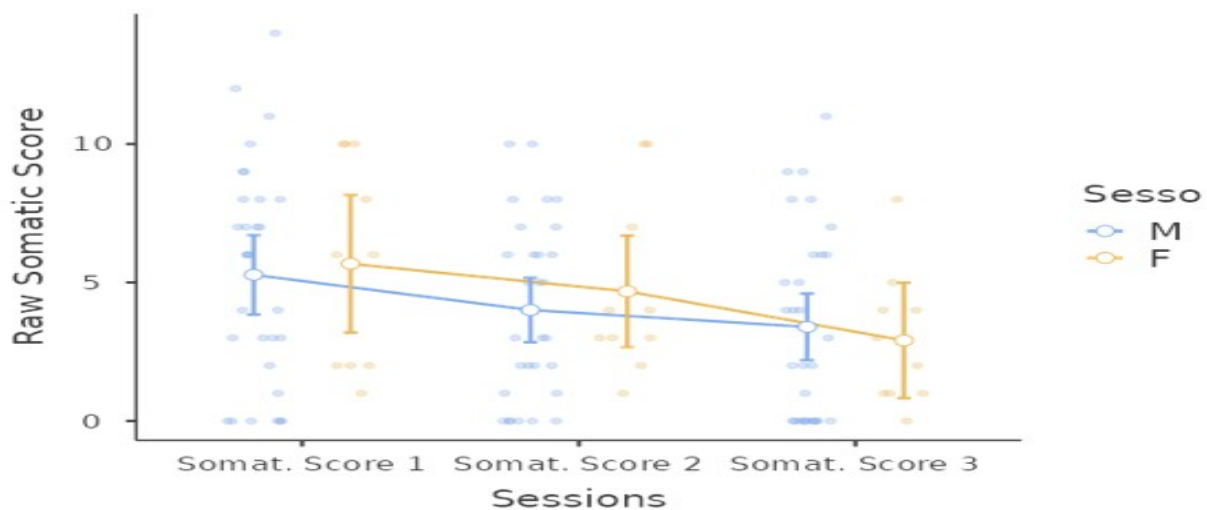

FIGURE 18

Figure 18 displays the trend over different sessions of the scores of the somatic factor questions, categorized by gender (males: M, females: F). The colored dots represent the raw values, the white circles describe the averages, and the vertical lines represent the standard deviations (SD).

| | Somma dei Quadrati | gdl | Media Quadratica | F | p | η²G |
|---|---|---|---|---|---|---|
| Sessions | 88.800 | 2 | 44.400 | 9.263 | < .001 | 0.061 |
| Residuo | 373.867 | 78 | 4.793 | | | |

FIGURE 19

The Figure 19 illustrates the results of the ANOVA test on the scores obtained from the somatic item questions, without considering the effect of other covariates. As can be observed from the "p value" (less than 0.001), the difference in scores obtained for the somatic item in the three subsequent sessions is statistically significant.

In Figure 20, the variations in scores obtained in the different sessions analyzed with the ANOVA test are reported, with the addition of non-significant covariates. By analyzing for the covariates "age" and "gender", statistical significance is reached for the covariates "sessions" and "age", while no significant trend is observed for the covariates "sessions" and "gender". This is evident by observing the values of the variances in the dataset. The increase in sample size and treatment time are factors that could have a significant impact on reaching statistical significance.

|  | Somma dei Quadrati | gdl | Media Quadratica | F | p | η²$_G$ |
|---|---|---|---|---|---|---|
| Sessions | 15.392 | 2 | 7.696 | 1.678 | 0.194 | 0.012 |
| Sessions ∗ Età | 28.534 | 2 | 14.267 | 3.111 | 0.050 | 0.021 |
| Sessions ∗ Sesso | 5.529 | 2 | 2.765 | 0.603 | 0.550 | 0.004 |
| Residuo | 339.377 | 74 | 4.586 |  |  |  |

FIGURE 20

The figure 20 describes the results of the ANOVA test on the scores of responses related to the somatic item, assuming that all cases were analyzed for sessions, age and gender covariates. The table indicate that the difference in scores calculated in the questionnaires across sessions is not statistically significant when considering sessions as a whole, but it is significant when considering sessions and age covariate.

In the next step of the study, we aimed to explore the correlations between the scores obtained in the questionnaire responses and the variable time. The objective is to investigate the existence and type of correlation between the performance of physical activity, evaluated in terms of depression levels, and the observation time among the three sessions. Our design was to extract information from the data to determine if there is a significant link between physical activity and depression, and to evaluate how the levels of depression measured by the questionnaires can change over time through the various physical activity sessions. To this end, we studied the possible correlations between all the scores in the sessions in a correlation matrix (Figure 21).

|  | | Punteggio - Prima sessione | Punteggio - Seconda Sessione | Punteggio - Terza Sessione |
|---|---|---|---|---|
| Punteggio - Prima sessione | r di Pearson | — | | |
| | gdl | — | | |
| | valore p | — | | |
| | Rho di Spearman | — | | |
| | gdl | — | | |
| | valore p | — | | |
| Punteggio - Seconda Sessione | r di Pearson | 0.824*** | — | |
| | gdl | 38 | — | |
| | valore p | <.001 | — | |
| | Rho di Spearman | 0.856*** | — | |
| | gdl | 38 | — | |
| | valore p | <.001 | — | |
| Punteggio - Terza Sessione | r di Pearson | 0.502*** | 0.579*** | — |
| | gdl | 38 | 38 | — |
| | valore p | <.001 | <.001 | — |
| | Rho di Spearman | 0.509*** | 0.551*** | — |
| | gdl | 38 | 38 | — |
| | valore p | <.001 | <.001 | — |

Nota. * p < .05, ** p < .01, *** p < .001

FIGURE 21

The figure 21 displays the results of the correlation matrix among the questionnaire scores and the physical activity sessions at increasing times. All correlations are strongly positive between the scores of the three items (affective, cognitive, and somatic) in the subsequent sessions (Pearson's R and Spearman's Rho coefficients are all more than 0.5), and all are statistically significant with p<0.001.

The correlation matrix demonstrates positive correlation coefficients between the scores of the questionnaires and the physical activity sessions at increasing times. The correlations are all positive between the scores of the three items (affective, cognitive, and somatic) in the subsequent sessions, and all are statistically significant with p<0.001.
This suggests that there is a strong positive correlation between the scores and the physical activity sessions, indicating that the physical activity sessions have a significant impact on the scores.
The possible reading of this data could be related to the maintenance or reduction over time of the scores obtained from the participants' questionnaire responses. This interpretation could indicate that, based on the collected data, there is a connection between the temporal trend of physical activity sessions and the levels of depression measured through the questionnaires. It is important to note that correlation does not imply causation, and further analysis is needed to determine if the physical activity sessions are causing the changes in the scores or if there is another underlying factor causing both the physical activity sessions and the changes in the scores. Similarly, we repeated the same analysis to find the possible correlations between the symptoms detected by the questionnaires before, during, and after the "training" in the various sessions in the correlation matrix (Figure 22).

|  | | Sintomi - Prima sessione | Sintomi - Seconda Sessione | Sintomi - Terza Sessione |
|---|---|---|---|---|
| Sintomi - Prima sessione | r di Pearson | — | | |
| | gdl | — | | |
| | valore p | — | | |
| | Rho di Spearman | — | | |
| | gdl | — | | |
| | valore p | — | | |
| Sintomi - Seconda Sessione | r di Pearson | NaN [a] | — | |
| | gdl | 37 | — | |
| | valore p | NaN | — | |
| | Rho di Spearman | 0.570*** | — | |
| | gdl | 37 | — | |
| | valore p | < .001 | — | |
| Sintomi - Terza Sessione | r di Pearson | NaN [a] | NaN [a] | — |
| | gdl | 37 | 38 | — |
| | valore p | NaN | NaN | — |
| | Rho di Spearman | 0.629*** | 0.640*** | — |
| | gdl | 37 | 38 | — |
| | valore p | < .001 | < .001 | — |

Nota. * $p < .05$, ** $p < .01$, *** $p < .001$

FIGURE 22

The figure 22 shows the correlation matrix between the symptoms of depression evaluated at the beginning, during, and after the physical activity sessions through the questionnaires over time. The table highlights positive and strongly statistically significant correlations through the values of Spearman's Rho coefficient (all values of Spearman's Rho are greater than 0.57). The coefficients have a "p value" less than 0.001 between the symptoms in the three subsequent sessions. In this analysis, the Pearson correlation coefficient is not defined because the symptom variable analyzed is a categorical parameter.

In this correlations analysis we obtain Spearman's coefficients, as the analyzed variable is categorical. The table highlights statistically significant correlations ($p<0.001$) between the symptoms in the different sessions (Spearman's Rho values are greater than 0.57). This suggests that the observed correlations are probably not due to casual event, but rather reflect true relationships between the symptoms of depression in different physical activity sessions. The results confirm that there is a positive statistically significant relationship between the reduction of depression symptoms detected by the questionnaires and the duration of physical activity over time.

Discussion

This study investigated the effect of physical activity, specifically the Rock Steady Boxing program, on the intensity of depressive symptoms in PD. The results confirm that depression is one of the most impactful non-motor symptoms in PD, in line with recent studies (24-27). Our study used the Beck Depression Inventory-II (BDI-II) to assess depressive symptoms, which has been validated for PD diagnosis (22-29). The BDI-II evaluates cognitive, affective, and somatic aspects of depression (30-31). This study highlights the importance of physical exercise as an additional therapy for PD symptom management (Figures 7-11) (32). Our results show a significant trend towards reduced depressive symptoms, regarding in particular affective and cognitive items (Figures 8-9), with potential benefits for improving depressive symptoms and stabilizing motor symptoms. Interestingly, the benefits for affective and cognitive factors appear earlier than somatic factors, suggesting different timeframes for perceiving improvements in different aspects of depression (Figures 12, 15, 18). The analysis of BDI-II scores across three sessions revealed a decrease in high-scoring responses, indicating overall improvement in depressive symptomatology (Figures 4, 6, 7). The study also found significant differences in BDI-II scores among the different sessions (Figura 10), indicating positive effects of repeated sessions over time of RSB on depressive symptoms. The correlation analysis supports these findings. In particular we have demonstrated statistically significant positive correlations between obtained scores and sessions and significant positive correlations between symptoms across sessions (Figures 12, 15, 18). Our results suggest that covariates, as age and gender, might influence the results, with age having a significant effect on depression reduction (Figures 17, 20). The somatic factor may be strongly influenced by temporal or emotional factors, indicating the need for a different or additional focus in patient evaluation and treatment. It could be hypothesized that the planning and implementation of personalized training could also contribute to the improvement of the effects of RSB program on symptoms of depression in these subjects. The study concludes that physical activity, particularly RSB, may positively impact depressive symptomatology in PD patients.

## Conclusions

Further research is needed to fully understand the underlying mechanisms and identify the characteristics of individuals who may benefit most from such interventions. This study has several limitations that are important to highlight for proper interpretation of the results and to guide future research to address any gaps or uncertainties.

The first limitation is the sample size and representativeness. The study sample is primarily male, with limited information on PD diagnosis and depression status at diagnosis. The adapted exercise protocol, designed by a certified trainer, ensured safety and progression but could be optimized with more information. The small sample size (40 participants) may limit generalizability and transferability to other PD populations. The cohort's representativeness might be influenced by age and gender covariates distribution.

Missing questionnaire responses and exclusion of some participants may also be a limitation, affecting cohort representativeness. The completion rate of the questionnaire may raise concerns about possible selection bias.

The absence of a control group is a significant limitation, as it limits the ability to attribute observed effects specifically to physical activity (RSB). A control group would allow for more accurate comparisons with placebo effects or other interventions.

The adapted exercise protocol may also be a limitation, as the certification of the trainer may raise questions about intervention standardization and study replicability.

The duration of the intervention is also a limitation, as the observation time and lesson frequency are time restricted. These details may affect long-term effects, which are not yet known.

The interpretation of results, particularly the measurement of depression, may also be a limitation. Despite using the widely recognized Beck Depression Inventory-II (BDI-II), depression evaluation may be influenced by subjective factors such as individual perception and self-presentation.

Regarding covariates, several biases are present. The lack of control for medication therapy is a significant limitation, as it may influence results, as some participants may be on medication, contributing to observed effects. The analysis of covariates was conducted separately for age and gender, and the effect of sample partitioning may affect the significance of associations.

Finally, the small sample size prevented other types of classical and machine learning analyses that require large sample sizes.

Despite these limitations, the prospect of integrating exercise programs into the management of depression in PD is promising and could offer additional benefits to patients' and their families' quality of life.

## References

1. Balestrino R, Schapira AHV. Parkinson disease. Eur J Neurol. 2020;27(1):27-42. doi: 10.1111/ene.14108.

2. Kreitzer AC, Malenka RC. Striatal plasticity and basal ganglia circuit function. Neuron. 2008;60(4):543-54. doi: 10.1016/j.neuron.2008.11.005.
3. de Lau LM e Breteler MM, Epidemiology of Parkinson's disease. Lancet Neurol. 2006;5,(6): 525–35.
4. GBD 2016 Parkinson's Disease Collaborators. Global, regional, and national burden of Parkinson's disease, 1990-2016: a systematic analysis for the Global Burden of Disease Study 2016. Lancet Neurol. 2018;17(11):939-953. doi: 10.1016/S1474-4422(18)30295-3. Erratum in: Lancet Neurol. 2021;20(12):e7.
5. Lesage S e Brice A, Parkinson's disease: from monogenic forms to genetic susceptibility factors. Hum. Mol. Genet., 2009 vol. 18, R1,, pp. R48–59
6. Jia F, Fellner A, Kumar KR. Monogenic Parkinson's Disease: Genotype, Phenotype, Pathophysiology, and Genetic Testing. Genes. 2022;13(3):471. doi.org/10.3390/genes13030471.
7. Nichols WC, Pankratz N, Marek DK, Pauciulo MW, Elsaesser VE, Halter CA, Rudolph A, Wojcieszek J, Pfeiffer RF, Foroud T; Parkinson Study Group-PROGENI Investigators. Mutations in GBA are associated with familial Parkinson disease susceptibility and age at onset. Neurology. 2009;72(4):310-6. doi: 10.1212/01.wnl.0000327823.81237.dl
8. Pycock CJ, Marsden CD. Central deopaminergic receptor supersensitivity and its relevance to Parkinson's disease. J Neurol Sci. 1977;31(1):113-21. doi: 10.1016/0022-510x(77)90009-0. PMID: 833607.
9. Poewe, Werner, 'Global Scales to Stage Disability in PD: The Hoehn and Yahr Scale', in Cristina Sampaio, Christopher G. Goetz, and Anette Schrag (eds), Rating Scales in Parkinson's Disease: Clinical Practice and Research (2012; online edn, Oxford Academic, 1 Sept. 2013), https://doi.org/10.1093/med/9780199783106.003.0258, accessed 8 Apr. 2024.
10. Hoehn MM, Yahr MD. Parkinsonism: onset, progression, and mortality. 1967. Neurology. 2001;57(10 Suppl 3):S11-26. PMID: 11775596.
11. Goetz, C.G., Poewe, W., Rascol, O., Sampaio, C., Stebbins, G.T., Counsell, C., Giladi, N., Holloway, R.G., Moore, C.G., Wenning, G.K., Yahr, M.D. and Seidl, L. Movement Disorder Society Task Force report on the Hoehn and Yahr staging scale: Status and recommendations The Movement Disorder Society Task Force on rating scales for Parkinson's disease. Mov. Disord. 2004;19: 1020-1028. https://doi.org/10.1002/mds.20213.
12. Blandini F, Nappi G, Tassorelli C, Martignoni E. Functional changes of the basal ganglia circuitry in Parkinson's disease. Prog Neurobiol. 2000;62(1):63-88. doi: 10.1016/s0301-0082(99)00067-2. PMID: 10821982.
13. Young CB, Reddy V, Sonne J. Neuroanatomy, Basal Ganglia. 2023 Jul 24. In: StatPearls [Internet]. Treasure Island (FL): StatPearls Publishing; 2023. PMID: 30725826.
14. Findley LJ, Gresty MA, Halmagyi GM. Tremor, the cogwheel phenomenon and clonus in Parkinson's disease. J Neurol Neurosurg Psychiatry. 1981 Jun;44(6):534-46. doi: 10.1136/jnnp.44.6.534. PMID: 7276968; PMCID: PMC491035.
15. Hunker CJ, Abbs JH. Uniform frequency of parkinsonian resting tremor in the lips, jaw, tongue, and index finger. Mov Disord. 1990;5(1):71-7. doi: 10.1002/mds.870050117. PMID: 2296262.
16. Mulroy E, Erro R, Bhatia KP, Hallett M. Refining the clinical diagnosis of Parkinson's disease. Parkinsonism Relat Disord. 2024:106041. doi: 10.1016/j.parkreldis.2024.106041.
17. Jankovic J. Parkinson's disease: clinical features and diagnosis. J Neurol Neurosurg Psychiatry. 2008;79(4):368-76. doi: 10.1136/jnnp.2007.131045. PMID: 18344392.
18. Kim SD, Allen NE, Canning CG, Fung VSC. Parkinson disease. Handb Clin Neurol. 2018;159:173-193. doi: 10.1016/B978-0-444-63916-5.00011-2. PMID: 30482313.
19. Marsh L. Depression and Parkinson's disease: current knowledge. Curr Neurol Neurosci Rep. 2013;13(12):409. doi: 10.1007/s11910-013-0409-5. PMID: 24190780; PMCID: PMC4878671.

20. Ray S, Agarwal P. Depression and Anxiety in Parkinson Disease. Clin Geriatr Med. 2020;36(1):93-104. doi: 10.1016/j.cger.2019.09.012. Epub 2019 Sep 10. PMID: 31733705.
21. Mele B, Van S, Holroyd-Leduc J, Ismail Z, Pringsheim T, Goodarzi Z. Diagnosis, treatment and management of apathy in Parkinson's disease: a scoping review. BMJ Open. 2020;10(9):e037632. doi: 10.1136/bmjopen-2020-037632. PMID: 32907903; PMCID: PMC7482451.
22. Beck AT, Steer RA, Brown GK. BDI-II: Beck depression inventory. Pearson, 1996.
23. Beck AT, Steer R. Manual for revised Beck Depression Inventory. New York: Psychological Corporation; 1987.
24. Maggi G, D'Iorio A, Aiello EN, Poletti B, Ticozzi N, Silani V, Amboni M, Vitale C, Santangelo G. Correction to: Psychometrics and diagnostics of the Italian version of the Beck Depression Inventory-II (BDI-II) in Parkinson's disease. Neurol Sci. 2023;44(7):2631. doi: 10.1007/s10072-023-06746-4. Erratum for: Neurol Sci. 2023;44(5):1607-1612. PMID: 36939947; PMCID: PMC10257586.
25. Beck, A. T., Steer, R. A., Brown, G. K., & Van der Does, A. J. W. (2002). BDI-II-NL handleiding [BDI-II-Dutch Manual]. The Netherlands: Psychological Corporation
26. Maggi, G., D'Iorio, A., Aiello, E.N. et al. Psychometrics and diagnostics of the Italian version of the Beck Depression Inventory-II (BDI-II) in Parkinson's disease. Neurol Sci 44, 2631, 2023.
27. DI-II Beck Depression Inventory – II Author: Aaron T. Beck, Robert A. Steer e Gregory K. Brown. Psychological Corp.; Harcourt Brace, San Antonio, Tex., Boston, cop. 1996.
28. https://www.jamovi.org Consultato a Dicembre 2023.
29. Jankovic J. Parkinson's disease: clinical features and diagnosis. J Neurol Neurosurg Psychiatry. 2008;79(4):368–376.
30. Leentjens AF, Verhey FR, Luijckx GJ, Troost J. The validity of the Beck Depression Inventory as a screening and diagnostic instrument for depression in patients with Parkinson's disease. Mov Disord. 2000, 6:1221-4.
31. Shu H-F, Yang T, Yu S-X, et al. Aerobic exercise for Parkinson's disease: a systematic review and Meta-analysis of randomized controlled trials. PLoS One. 2014;9(7):e100503.
32. Sangarapillai K, Norman BM, Almeida QJ. Boxing vs Sensory Exercise for Parkinson's Disease: A Double-Blinded Randomized Controlled Trial. Neurorehabilitation and Neural Repair. 2021;35(9):769-777.